\documentclass[10pt,twocolumn,superscriptaddress,prl,showpacs,floatfix,aps]{revtex4-2}
\usepackage[utf8]{inputenc}
\usepackage{amsmath}
\usepackage{amssymb}
\usepackage{graphicx}
\usepackage{svg}
\usepackage{xspace}
\usepackage{mathtools}
\usepackage{amsfonts}
\usepackage{bm}
\usepackage{bbm}
\usepackage{dsfont}
\usepackage{soul}
\usepackage{comment}
\usepackage{physics}
\usepackage[normalem]{ulem}
\usepackage{babel}
\usepackage{xcolor}

\usepackage[final]{hyperref} 
\hypersetup{
	colorlinks=true,    
	linkcolor=blue,    
	citecolor=blue,    
	filecolor=magenta,   
	urlcolor=blue     
}

\begin{document}

\title{Exponential Speedup of Entanglement Generation by Quantum Mpemba Effects}

\author{Sara M. Benjadi}
\affiliation{Institut f\"ur Theoretische Physik, Heinrich-Heine-Universit\"at, D-40225 D\"usseldorf, Germany}

\author{Reinhold Egger}
\affiliation{Institut f\"ur Theoretische Physik, Heinrich-Heine-Universit\"at, D-40225 D\"usseldorf, Germany}

\author{Igor Gornyi}
\affiliation{\mbox{Institute for Quantum Materials and Technologies, Karlsruhe Institute of Technology, 76021 Karlsruhe, Germany} }
\affiliation{\mbox{Institut f\"ur Theorie der Kondensierten Materie, Karlsruhe Institute of Technology, 76128 Karlsruhe, Germany}}
\author{Andrea Nava}
\affiliation{Institut f\"ur Theoretische Physik, Heinrich-Heine-Universit\"at, D-40225 D\"usseldorf, Germany}

\begin{abstract}
Entanglement is a key resource for quantum technologies. We show that protocols employing quantum Mpemba effects allow one to exponentially accelerate the generation of entanglement, or to slow down the decay thereof. Two entanglement Mpemba effects with different operational meaning are introduced, focusing either on the task of rapidly generating a certain threshold value for entanglement or on achieving the asymptotic steady-state value. We show that entanglement Mpemba effects depend on the chosen entanglement measure. Using cluster elimination methods, many-body quantum systems also benefit from the exponential speedup of entanglement generation, as we demonstrate for a dissipative long-range Ising chain.
 \end{abstract}
\maketitle

\emph{Introduction.---}Entanglement \cite{Horodecki_2009} is a resource of central importance for many quantum information processing tasks, e.g., quantum communication and computation \cite{Nielsen_Chuang_2010}, teleportation \cite{Bennett_1993}, metrology \cite{Dooley_2023,Shen_2025}, cryptography \cite{Ekert_1991}, or error correction \cite{Brun_2006,Terhal2015,Bravyi_2025}. Consequently, a central challenge is to generate, preserve, or remove entanglement as efficiently as possible \cite{Vu2012,Liu_2021,Pauwels_2022,Eckart_2023}. In this Letter, we show that exponential speedups in the dynamical generation of entanglement, or alternatively in the slowdown of entanglement decay, can be achieved by employing quantum Mpemba effects (MEs). 
MEs refer to anomalous nonequilibrium phenomena after a parameter quench where, e.g., a hot initial state relaxes faster than a cold initial state toward the same (very cold) stationary state. 
For recent reviews, see Refs.~\cite{Teza2025,Ares2025}. MEs are presently under intense study both in classical \cite{Mpemba1969,Lu2017,Lasanta2017,Klich2019,Chetrite2021,Ibanez2024,shapira2026a} and in quantum systems \cite{Nava2019,Rylands2023,Nava2024,Murciano_2024,Liu2024,Moroder2024,Ares2025b,Zatsarynna2025,Su2025,Wang2024,Giulio2025,Strachan2024,Turkeshi2024,Qian2025,Lacerda2025,Westhoff2025,Nava2025,Nava_2025,Peluso_2026,Li2026a,Chattopadhyay2026,Melles_2026,Summer_2026,Bagui2026,chang2026,Xu2026a,Liu2026}. They can largely be understood from the spectral structure of the dynamical generator: Relaxation is governed by a hierarchy of modes, where for certain initial states, the overlap with the slowest modes is suppressed and the system evolves through fast decay channels only. 
For experimental reports of exponentially accelerated relaxation due to MEs, see, e.g., Refs.~\cite{Joshi2024,Shapira2024,Zhang2025,Tian_2025,Xu_2026}.

Existing theories typically characterize MEs through the dynamics of
distance-from-equilibrium functions such as the trace distance or the entanglement asymmetry 
\cite{Teza2025,Ares2025}. A fruitful alternative viewpoint \cite{Summer_2026} is to ask how rapidly quantum resources \cite{Chitambar_2019} are generated or consumed. We here introduce \emph{entanglement MEs} by studying the time evolution of an entanglement monotone $E(\rho)$ \cite{Vidal_2000}, where $\rho$ is the system state, see also Ref.~\cite{Aolita_2015}.
Different monotones probe distinct features of the state and capture inequivalent aspects of quantum correlations \cite{Vidal_2002,Acin_2012,Basso_2022,Kvorning_2022}. For mixed states, examples of entanglement monotones are logarithmic negativity and concurrence. For globally pure states, the entanglement entropy and R\'enyi entropies provide further entanglement measures. Our theory applies to generic choices for $E(\rho)$, but different choices result in different relaxation times and entanglement MEs.
This degree of nonuniversality is typical for Mpemba physics \cite{Teza2025,Ares2025,Summer_2026}.
We next introduce two complementary operational definitions. 

We first define the \emph{preparation entanglement ME} (PEME), aimed at the fast preparation of highly entangled states from low-entangled initial states.
Consider two weakly entangled initial states evolving toward the same highly entangled stationary state $\rho^{\rm (st)}$ with entanglement $E_\infty=E(\rho^{\rm (st)})$, and 
let $E_{\rm tar}<E_\infty$ be the actual target value needed for a given task \cite{Brun_2006,Bravyi_2025,Wang_2025}. In applications, approaching the ideal value $E_\infty$ is often less important than rapidly generating entanglement above a threshold $E_{\rm tar}$. With $E(t)=E(\rho(t))$,
the entanglement preparation time $\tau_{\rm prep}(E_{\rm tar})$ is defined as the earliest time such that $E(t)\ge E_{\rm tar}$ for all $t\ge \tau_{\rm prep}$. 
The PEME occurs if the state $\rho^{\rm (L)}$ with initially lower entanglement reaches and permanently surpasses $E_{\rm tar}$ faster than the state $\rho^{\rm (H)}$ with initially higher entanglement, 
i.e., if $\tau_{\rm prep}^{\rm (L)}<\tau_{\rm prep}^{\rm (H)}$ for $E^{\rm (L)}(0)<E^{\rm (H)}(0)$. 
Since in contrast to conventional MEs \cite{Teza2025}, this criterion is not based on the parity of transient crossings between $E^{\rm (L)}(t)$ and $E^{\rm (H)}(t)$, the PEME is robust against oscillations or revivals. In fact, $\tau_{\rm prep}$ is determined by the overlap of the initial state with all spectral modes (see below), not only with the slowest ones.  

We next introduce the \emph{relaxation entanglement ME} (REME), characterizing the asymptotic approach of $E(t)$ to $E_\infty$. Since the stationary state is only reached for $t\to \infty$, one has to specify a threshold accuracy ($E_{\rm th}$) which physically corresponds to the experimental resolution limit. The relaxation time $\tau_{\mathrm{rel}}(E_{\mathrm{th}})$ is defined as the final-entry time, such that $|E(t)-E_\infty|\le E_{\rm th}$ for all $t\ge \tau_{\rm rel}$. The REME then occurs for $\tau_{\rm rel}^{\rm (L)}<\tau_{\rm rel}^{\rm (H)}$ when comparing two weakly entangled initial states.
Both PEME and REME differ from conventional MEs defined, say, through the trace distance: Two initial states may show the PEME and/or REME without a conventional ME, and vice versa. In fact, the parity of crossings between both $E(t)$ curves is not correlated with entanglement MEs, and for the REME, $E(t)$ can converge to $E_\infty$ 
from above or below. Furthermore, $\tau_{\rm prep}$ (and thus the PEME) depends on the target value $E_{\rm tar}$, while the ordering $\tau_{\rm rel}^{\text{(L)}}<\tau_{\rm rel}^{\text{(H)}}$ (and thus the REME) becomes independent of $E_{\rm th}$ for sufficiently small $E_{\rm th}$. 

Similarly, inverse entanglement MEs can be defined if entanglement is progressively lost, e.g., due to decoherence. Consider two highly entangled initial states evolving toward the same weakly entangled stationary state with small $E_\infty$. The inverse entanglement ME occurs if, compared to a less entangled initial state (L), an initially more entangled state (H) first reaches and subsequently remains below a small threshold value $E_{\rm tar}$ (for the PEME),
or first reaches the final value $E_\infty$ within accuracy $E_{\rm th}$ (for the REME). In practice, initial states that suppress inverse entanglement MEs are useful in slowing down entanglement decay. This perspective is especially relevant for quantum memories and quantum communication \cite{Liu_2024}, where preservation of entanglement remains a key challenge \cite{Zhou_2022}. We emphasize that PEME and REME probe distinct dynamical aspects: 
the PEME refers to the rapid preparation of a useful quantum resource, while the REME characterizes the long-time relaxation toward $E_\infty$. Building on 
Refs.~\cite{Carollo_2021,Kochsiek2022,Bao2025,Caldas_2026,Beato_2026}, we also show below that slow modes controlling the entanglement dynamics can be systematically eliminated through unitary rotations of the initial state. 
This approach exponentially accelerates the preparation and manipulation of entanglement and is of key importance for many-body quantum systems \cite{Fazio_2025}, as we illustrate for a dissipative long-range Ising chain.

\emph{Spectral theory of entanglement MEs.---}We assume that the mixed state $\rho(t)$ of an open quantum system with Hamiltonian $H$ evolves according to the Lindblad master equation \cite{Lindblad1976,breuer2007theory},  
\begin{equation}
\dot{\rho} = \mathcal{L}[\rho] = -i[H,\rho] + \sum_\mu \left( K_\mu\rho K_\mu^\dagger - \frac12 
\left\{ K_\mu^\dagger K^{}_\mu,\rho \right\} \right) \label{eq:lindblad}
\end{equation}
with jump operators $\{ K_\mu\}$ describing weak couplings to a Markovian environment. Assuming a unique stationary state $\rho^{\rm (st)}$, we use a spectral decomposition of the superoperator ${\cal L}$. Denoting by $R_n$ ($L_n$) its right (left) eigenoperators, $\mathcal{L}[R_n]=\lambda_n R_n$ ($\mathcal{L}^\dagger[L_n]=\lambda^*_n L_n$), and ordering the eigenvalues $\{\lambda_n\}$ as $0=\lambda_0\ge {\rm Re}\lambda_1 \ge {\Re}\lambda_2\ge \cdots,$ one finds
\begin{equation} 
\rho(t) = \rho^{\rm (st)} + \sum_{n\ge1} c_n e^{\lambda_n t} R_n, \quad c_n = \mathrm{Tr}\left(L_n^\dagger \rho(0)\right),
\label{eq:rho_expansion} 
\end{equation}
with $R_0=\rho^{\rm (st)}$, $L_0=\mathbbm{1}$, and ${\rm Tr}(L^\dagger_n R_m)=\delta_{nm}$. At long times, the dynamics is dominated by the lowest modes. 
Comparing Eq.~\eqref{eq:rho_expansion} for two initial states $\rho^{\rm (L,H)}(0)$, the strong (conventional) ME occurs if $c^{\rm (L)}_1=0$ but $c_1^{\rm (H)}\ne 0$ such that the initially farther state (L) is controlled by the faster mode ($\lambda_2$) \cite{Lu2017,Carollo_2021}. 
Let us now consider an arbitrary entanglement monotone $E(\rho)$. Assuming Fréchet differentiability at $\rho=\rho^{\rm (st)}$ and writing $\delta\rho(t)=\rho(t)-\rho^{\rm (st)}$, we expand $E(\rho)$ near the stationary state, 
\begin{equation} \label{expansion}
E(\rho) = E_\infty + \left. \frac{\delta E}{\delta\rho} \right|_{\rho^{\rm (st)}}\!\!\! 
[\delta\rho] + \frac12  \left. \frac{\delta^2 E}{\delta\rho^2} \right|_{\rho^{\rm (st)}}\!\!\! [\delta\rho,\delta\rho] 
+ {\cal O}(\delta\rho^3).
\end{equation} 
Inserting Eq.~\eqref{eq:rho_expansion} into Eq.~\eqref{expansion} yields 
\begin{eqnarray} \nonumber
E(t) &\simeq &E_\infty + \sum_{n\ge 1} c_n \mathcal{Z}_n e^{\lambda_n t}\! +\!\! \sum_{n,m\ge 1} c_n c_m {\mathcal Z}_{nm} e^{(\lambda_n+\lambda_m)t}
, \\ \label{eq:E_expansion} 
\mathcal{Z}_n &=& \left. \frac{\delta E}{\delta\rho} \right|_{\rho^{\rm (st)}}\!\!\! [R_n] ,\quad
\mathcal{Z}_{nm} = \frac12 \left. \frac{\delta^2 E}{\delta\rho^2} \right|_{\rho^{\rm (st)}} \!\!\![R_n,R_m],
\end{eqnarray}
where $\mathcal{Z}_n$ ($\mathcal{Z}_{nm}$) is the first-order (second-order) Fréchet derivative, i.e, linear-response functional, of $E(\rho)$ at $\rho^{\rm (st)}$ evaluated on $R_n$ (and $R_m$). 
Importantly, the (generally complex-valued) numbers ${\cal Z}_n$ and ${\cal Z}_{nm}$ do not depend on the initial state. Corrections to Eq.~\eqref{eq:E_expansion} scale as ${\cal O}\left(e^{3{\rm Re}\lambda_1 t}\right)$ and are due to 
$\delta \rho^3$-terms in Eq.~\eqref{expansion}, yielding third-order coefficients ${\cal Z}_{nmk}$. The long-time dynamics of $E(t)$ is thus governed by the same spectral modes as $\rho(t)$ with the amplitudes $c_n$ in Eq.~\eqref{eq:rho_expansion} renormalized by $\mathcal{Z}_n$, and by additional modes 
due to nonlinearities governed by combinations of eigenvalues of the Lindbladian. These appear in Eq.~\eqref{eq:E_expansion}, e.g., through the ${\cal Z}_{nm}$ terms.
Different monotones $E(\rho)$ and/or system bipartitions are characterized by different ${\cal Z}$ coefficients and thus need not exhibit identical entanglement MEs. 

\begin{figure}
  \centering
  \includegraphics[width=1\linewidth]{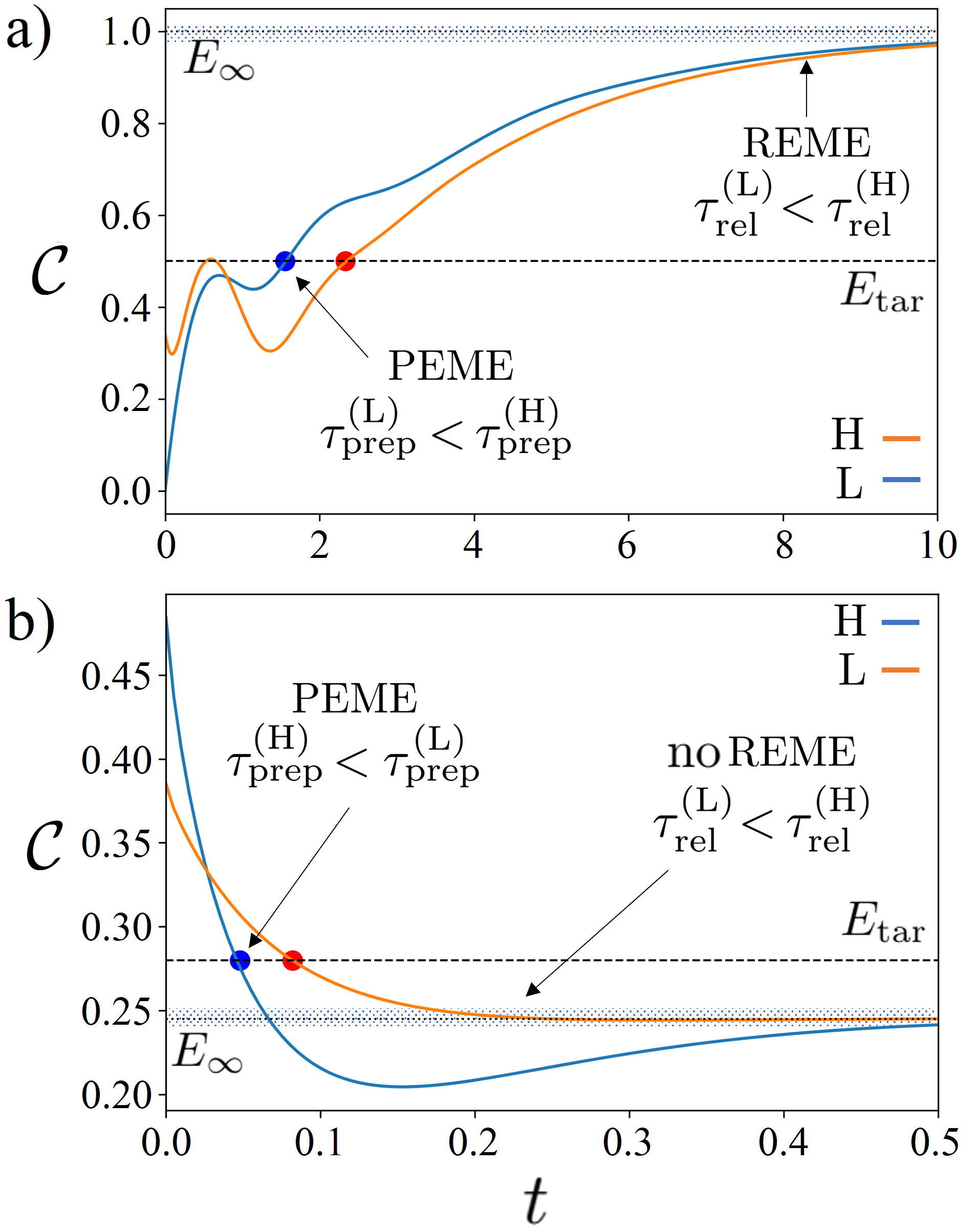}
  \caption{Entanglement MEs for the concurrence, $E(\rho)={\cal C}(\rho)$, of the I2KM, see Eq.~\eqref{eq}, with $J_h=1$ and other parameters specified in the End Matter. The shown results were obtained by numerically solving Eq.~\eqref{eq:lindblad}.
  (a) Direct effect: Two weakly entangled initial states (L, H) evolve toward a highly entangled stationary state with ${\cal C}(\infty)=E_\infty$. The dotted region around $E_\infty$ has the width $2E_{\rm th}$.
  Since ${\cal C}^{{\rm (L)}}(t)$ (blue curve) reaches and permanently surpasses the threshold $E_\mathrm{tar}$ earlier than ${\cal C}^{\rm (H)}(t)$ (orange curve), the PEME occurs. We also observe the REME at long times. 
  (b) Inverse effect: Two highly entangled initial states (H, L) approach a weakly entangled stationary state for $t\to \infty$. 
  The inverse PEME is observed as earlier threshold ($E_{\rm tar}$) crossing for ${\cal C}^{\rm (H)}(t)$ (blue). However, the inverse REME is absent, showing its independence of the PEME.}
  \label{fig:fig1}
\end{figure}

Equation \eqref{eq:E_expansion} then determines both the REME and the PEME. For simplicity, let us assume 
real-valued and non-degenerate eigenvalues $\{\lambda_n\}$ in what follows; for the general case, see the End Matter.
First, with $E^{\rm (L)}(0)<E^{\rm (H)}(0)$, the condition
$|E^{\rm (L)}(t)-E_\infty|<|E^{\rm (H)}(t)-E_\infty|$ at large $t$ implies that the REME occurs for
\begin{equation}\label{REMEcond}
  \left| c_1^{\rm (L)} {\cal Z}_1 \right| < \left| c_1^{\rm (H)} {\cal Z}_1 \right|.
\end{equation}
For ${\cal Z}_1=0$, an important difference to conventional MEs arises since the linear-in-$\rho$ contribution to $E(t)$ is now blind to the slowest mode ($\lambda_1$) for \emph{all}\ initial states. 
Since generically ${\cal Z}_{11}\ne 0$, one expects from Eq.~\eqref{eq:E_expansion}
the long-time behavior $E(t)-E_\infty \propto e^{-\Gamma t}$ with the decay rate
$\Gamma={\rm min}(2|{\rm Re}\lambda_1|, |{\rm Re}\lambda_n|)$, where $n>1$ is the
smallest index with $c_n {\cal Z}_n\ne 0$. A first mechanism for exponential entanglement speedup is to tune ${\cal Z}_k=0$ for all $k<n$, so that the entanglement decays at $\Gamma=\min(2|\mathrm{Re}\lambda_1|,|\mathrm{Re}\lambda_n|)$. Since the nonlinear term at $2|\mathrm{Re}\lambda_1|$ survives whenever $c_1\neq0$, this yields a speedup of at most a factor of two. The condition $\mathcal{Z}_k = 0$ is robust (in fact, symmetry-protected \cite{unpublished}) and does not rely on fine-tuning of Lindbladian parameters (see End Matter). A second mechanism is to impose $c_k=0$ for all $k\le n$, which removes those modes from $\rho(t)$ and hence also from the corresponding nonlinear terms in Eq.~\eqref{eq:E_expansion}. The decay rate is then $\Gamma=\min(2|\mathrm{Re}\lambda_{n+1}|,|\mathrm{Re}\lambda_m|)$ with $m>n$ the smallest index with ${\cal Z}_m\neq0$; it equals $|\mathrm{Re}\lambda_{n+1}|$ if ${\cal Z}_{n+1}\neq0$, in which case iterating the scheme gives arbitrarily large speedups.
For the PEME, a similar but different condition follows by imposing $E^{{\rm (H)}}(t)<E^{{\rm (L)}}(t)$ at long times,
\begin{equation}\label{PEMEcond}
  \left (c^{{\rm (H)}}_1-c^{{\rm (L)}}_1\right) \mathcal{Z}_1<0.
\end{equation} 
For $\mathcal{Z}_1 = 0$, one again needs to consider the first $c_n{\cal Z}_n\ne 0$ term and compare it to the ${\cal Z}_{11}$ contribution. Since the PEME is concerned with short-to-intermediate times, Eq.~\eqref{PEMEcond} is approximate, and in general a numerical analysis of Eq.~\eqref{eq:E_expansion} is needed to validate this criterion.
Similar arguments hold for the inverse effects. The above discussion shows that the REME and/or PEME may arise without conventional MEs, and vice versa. Equations \eqref{REMEcond} and \eqref{PEMEcond} suggest that REME and PEME are independent phenomena for ${\rm sgn}\left(c_1^{\rm (L)} {\cal Z}_1\right) \ne {\rm sgn}\left(c_1^{\rm (H)} {\cal Z}_1\right)$. An example for this independence is shown in Fig.~\ref{fig:fig1}(b).

\emph{Interacting two-site Kitaev model (I2KM).---}As simple yet nontrivial first example, we consider the dissipative I2KM \cite{Leijnse_2012,Tsintzis_2024,Samuelson_2024}, which is experimentally accessible in proximitized double quantum dots and allows for engineering poor man's Majorana bound states \cite{Dvir2023,Bordin_2024,tenHaaf2024}. We have spinless fermions on two sites with energies $\epsilon_{1,2}$, coupled by a tunneling amplitude $J_h$, a superconducting pairing amplitude $\Delta$, and the Coulomb interaction $U$. The Hamiltonian is
\begin{equation}
H=\sum_{i=1,2}\epsilon_i n_i +U n_1n_2+\left(J_h c_1^\dagger c_2+ \Delta c_1^\dagger c_2^\dagger + {\rm h.c.}\right),
\label{eq}
\end{equation}
where $c_i$ annihilates a fermion on site $i$ and $n_i=c_i^\dagger c_i^{}$. The I2KM is
coupled to an environment through a set of jump operators, see Eq.~\eqref{eq:lindblad}. These operators and our
parameter choices are specified in the End Matter.
 For this model, which is equivalent to two coupled dissipative spin-$\frac12$ systems, we employ the concurrence $\mathcal{C}(\rho)$ as entanglement monotone, ranging from ${\cal C}=0$ (separable) to ${\cal C}=1$ (maximally entangled states) \cite{Wootters1998}. In Fig.~\ref{fig:fig1}, we illustrate the PEME and the REME for the I2KM, showing that both effects can be realized. 

We next note that for nearly identical Lindbladian parameters and the same initial states, essentially indistinguishable trace-distance curves, ${\cal D}_T(t)=\frac12 {\rm Tr}|\rho(t)-\rho^{\rm (st)}|$,
are expected. Even in such cases, Eq.~\eqref{eq:E_expansion} suggests that entanglement MEs may occur due to the spectral filtering encoded by the ${\cal Z}$ coefficients. However, such effects are sensitive to the choice for $E(\rho)$. For a bipartite system $A \cup B$, one can use the logarithmic negativity 
$\mathcal{N}(\rho) = \log_2\|\rho^{T_B}\|_1$ as entanglement monotone,
where $T_B$ is the partial transpose with respect to subsystem $B$ and $\|\cdot\|_1$ the trace norm \cite{Vidal_2002}. 
If $\rho^{{\rm (st)}\, T_B}$ has no zero or degenerate negative eigenvalues, the Fréchet differential is well defined and gives 
\begin{equation}\label{eq:ZN}
  \mathcal{Z}^{(\mathcal{N})}_n = -\frac{2}{\| \rho^{ \mathrm{(st)}\, T_B} \|_1 \ln{2}} \sum_j \langle \phi_j | R_n^{T_B} | \phi_j \rangle,
\end{equation}
where $|\phi_j\rangle$ are eigenvectors to negative eigenvalues of $\rho^{\mathrm{(st)}\,T_B}$. 
By comparing entanglement MEs for the I2KM based on either ${\cal C}(\rho)$ or ${\cal N}(\rho)$, rather different 
results can emerge as we demonstrate in the End Matter. 

\emph{Many-body systems.---}We now turn to many-body systems, where $\mathcal{L}$ has an exponentially large number of eigenvalues $\{\lambda_n\}$. Rather than being isolated, they are typically organized in clusters with similar decay rates and comparable spatial structure of the corresponding eigenmodes \cite{Macieszcak2016,Rose_2016,Macieszczak_2021,Matern_2023}. The long-time relaxation of $\rho(t)$ is then governed by the slowest spectral cluster $C_1$, formed by a bundle of several slow modes with ${\rm Re}(\lambda_n)\simeq {\rm Re}(\lambda_1)$. Similarly, higher-order clusters $C_{m>1}$ correspond to progressively faster relaxation sectors.
The long-time dynamics is then described by
$\rho(t)=\rho^\mathrm{(st)}+\sum_{n \in C_1} c_n e^{\lambda_n t} R_n
+\sum_{n \in C_2} c_n e^{\lambda_n t} R_n + \cdots,$
with related expressions for $E(t)$, see Eq.~\eqref{eq:E_expansion}.
In order to achieve exponential acceleration in MEs, instead of eliminating the slowest eigenmode, one may suppress the entire slow-mode cluster by imposing $c_n=0$ for $n\in C_1$. The corresponding operation can be realized by an experimentally accessible \cite{Caneva2009,Ansel_2024} unitary transformation of the initial state \cite{Beato_2026}, 
  $\rho_\perp(0) = U \rho(0) U^\dagger.$ Note that this initial-state rotation itself can already change entanglement. However, the stationary state $\rho^{\rm (st)}$ is generally a mixed state.
In many-body systems, the condition for exponential speedup of entanglement follows as 
$c_n= 0$ for all $n\in C_1$. 
This cluster elimination can be implemented by state preparation and/or the above unitary rotation. The next spectral sector $C_2$ then takes over and leads to exponentially faster entanglement generation. 
In the example below, we have followed the numerical approach of Ref.~\cite{Beato_2026}
to implement the unitary, but in principle also dissipative operations can be used.
One may now iterate the scheme and eliminate even more clusters. Alternatively, one may tune ${\cal Z}_n=0$ for all $n\in C_1$. As discussed above, however, this at best results in an entanglement speedup by a factor two.

\emph{Long-range Ising model.---}As example, we consider the dissipative long-range Ising model \cite{Koffel_2012} with open boundary conditions and Hamiltonian 
\begin{equation}\label{ising}
  H = h_x \sum_{i=1}^{N} \sigma_i^x + \sum_{i<j} \frac{J}{|i-j|^\alpha} \sigma_i^z \sigma_j^z,
\end{equation} 
where $N$ is the chain length, $\sigma_i^{x,z}$ are spin-$\frac12$ Pauli operators at site $i$, $h_x$ is
a transverse field strength, $J$ sets the interaction scale, and $\alpha$ controls the interaction range. 
For $\alpha < 1$, interactions are effectively long-range and cause collective correlations. 
We employ dissipation engineering which tends to drive the system to the vicinity of a GHZ state, 
$|{\rm GHZ}\rangle =(|00\cdots 0\rangle+|11\cdots1\rangle)/\sqrt{2}$ \cite{Greenberger_1990}, by means of the local jump operators $K_i = \frac12\sqrt{\gamma} \,\sigma_{i+1}^{x} (1- \sigma_i^z\sigma_{i+1}^z)$ (with $i=1,\dots,N-1$) and a global jump operator $K_g = \frac12\sqrt{\gamma} \,\sigma_1^z (1- \prod_{i=1}^{N}\sigma_i^x)$. 
The $\sigma_i^z\sigma_{i+1}^z$ stabilizers in $K_i$ prefer spin alignment between neighboring sites, 
while the global stabilizer in $K_g$ locks the relative phase factor between 
$|00\cdots0\rangle$ and $|11\cdots1\rangle$ \cite{Barreiro2011}.
For numerical results, we use the value $\alpha=0.5$ characteristic for trapped ion experiments \cite{Britton2012,Zhang2017} and consider $N=6$ sites with a $3|3$ bipartition, employing $E(\rho)={\cal N}(\rho)$.

\begin{figure}
  \centering
  \includegraphics[width=1\linewidth]{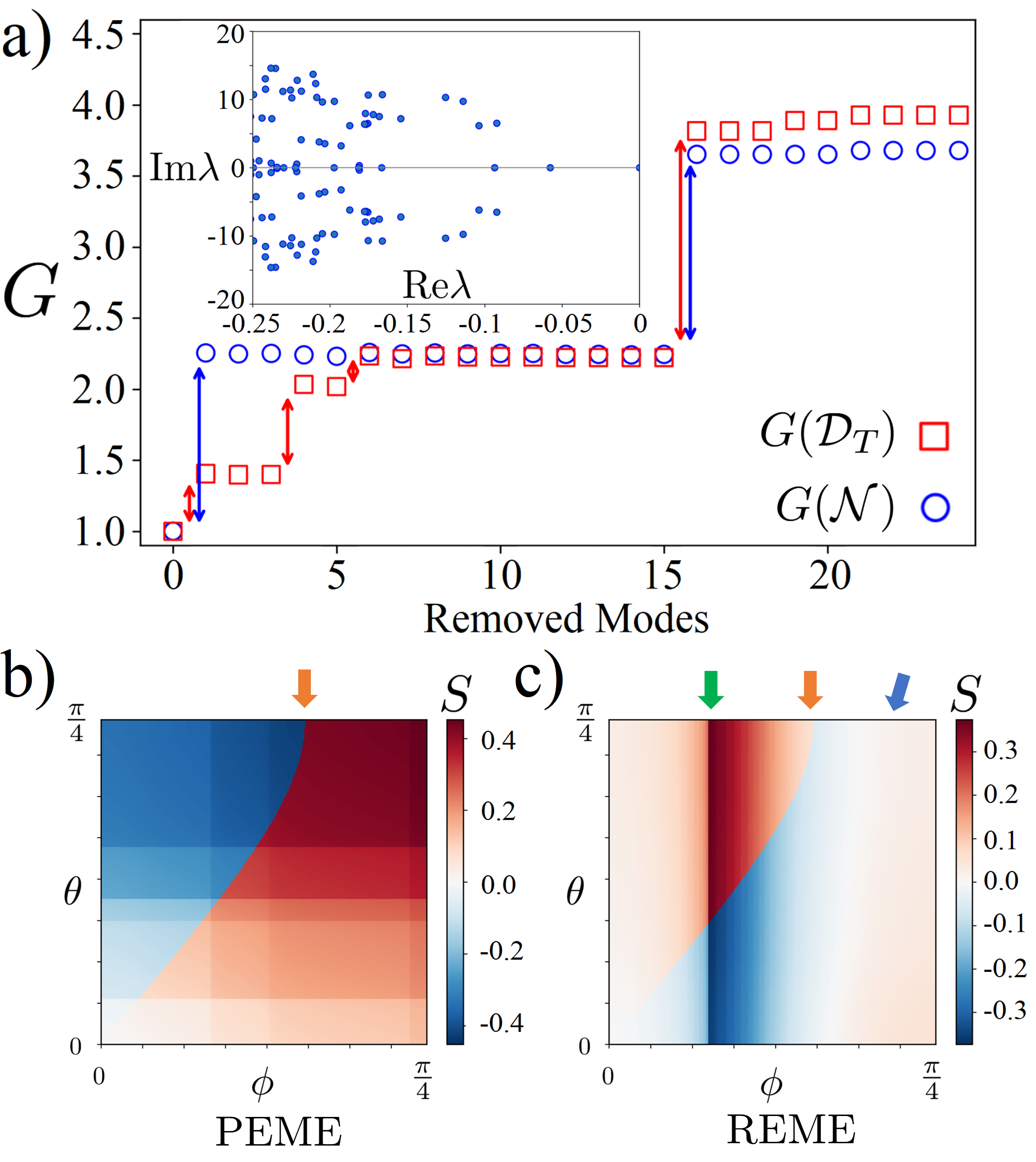}
  \caption{Results for the dissipative long-range Ising model with $N=6$ sites, see Eq.~\eqref{ising}, using $E(\rho)={\cal N}(\rho)$ with a $3|3$ bipartition. They are obtained from Eq.~\eqref{eq:lindblad} for 
  $J=1$, $\alpha= 0.5$, and $h_x=5/6$, using ${\cal D}_{T,{\rm th}}=10^{-3}$, ${\cal N}_{\rm th}=10^{-5}$.
  (a) Relaxation time speedups $G({\cal D}_T)$ and $G({\cal N})$, see Eq.~\eqref{gain}, vs number of removed Lindbladian modes, for $\gamma=1/6$. The inset shows the slow-mode part of the spectrum of ${\cal L}$. 
  (b) Normalized time difference $S$ for the PEME, defined as in Eq.~\eqref{gain} but with $\tau_{\rm rel}\to 
  \tau_{\rm prep}$, shown as color-scale plot in the $\phi$--$\theta$ plane. These angles parameterize the two initial states \eqref{initial}. The PEME corresponds to $S>0$. We use $\gamma=1/3$ and $E_{\rm tar}=0.8$. 
  (c) Same as (b) but for the REME. In (b) and (c),
  directions along which a relabeling discontinuity (continuous transition) is realized are indicated by
  orange (blue) arrows. Green arrows indicate directions with exponential speedup.
  }
  \label{fig:fig2}
\end{figure}

In Fig.~\ref{fig:fig2}, we show results for the relaxation time speedup $G$ and for the normalized time difference $S$ for initial states $\rho^{\rm (L,H)}(0)$, 
\begin{equation}\label{gain}
  G = \frac{\tau_\mathrm{rel}(\rho)}{\tau_\mathrm{rel}(\rho_\perp)}, \quad S=\frac{\tau^{\rm (H)}_{\mathrm{rel}}-\tau^{\rm (L)}_{\mathrm{rel}}}{\tau^{\rm (H)}_{\mathrm{rel}}+\tau^{\rm (L)}_{\mathrm{rel}}}.
\end{equation}
The definition of $S$ in Eq.~\eqref{gain} applies to the REME. For the PEME, one simply replaces
$\tau_{\rm rel}\to \tau_{\rm prep}$. First, in Fig.~\ref{fig:fig2}(a), starting from $\rho(0) = \ket{00\cdots 0}\!\bra{00\cdots 0}$, the robustness of the relaxation speedup due to cluster elimination
is shown both for the trace distance, $G({\cal D}_T)$, and for the REME using the logarithmic negativity, $G({\cal N})$, vs the number of removed Lindbladian modes. For instance, $G({\cal N})$ follows from Eq.~\eqref{gain} with $\tau_{\rm rel}$ determined from $|\mathcal{N}(t)-\mathcal{N_\infty}|\le \mathcal{N}_{\mathrm{th}}$ for $t\ge \tau_{\rm rel}$.
The precise value of $G({\cal D}_T)$ and/or $G({\cal N})$ depends on the threshold (${\cal D}_{T,{\rm th}}$ and ${\cal N}_{\rm th}$, respectively), and thus on the time $t$. What is of more interest is that both $G({\cal D}_T)$ and $G({\cal N})$ display a plateau structure separated by abrupt jumps, indicating that the $\{\lambda_n\}$ contribute in clusters $C_m$ rather than independently. Removing the slowest mode leads to a jump in both $G({\cal D}_T)$ and $G({\cal N})$. Here, the unitary implementing the mode elimination changes the entanglement content and affects the jump structure of $G({\cal N})$ observed in Fig.~\ref{fig:fig2}(a). We discuss this in more detail in the End Matter.
Remarkably, $G({\cal N})$ is then blind to the elimination of successive modes and no further jumps occur up to the $17$th mode, while two more jumps take place in $G({\cal D}_T)$.
This difference highlights that the entanglement dynamics follows from a spectrally filtered projection of the Lindbladian spectrum.
By eliminating just the lowest mode, one already has an exponential speedup $G({\cal N})\approx 2.3$, 
while eliminating the lowest 17 modes gives $G({\cal N})\approx 3.6$. 
We note that each removed mode imposes one constraint, while for a pure initial state in $d=2^N$ dimensions, the set $U\rho(0) U^\dagger$ has real dimension $2d-2$. A dimension count is consistent with a large family of solutions, provided the corresponding real constraints are independent and regular, which explains why cluster elimination works.
We note that we found \cite{unpublished} similar behavior as in Fig.~\ref{fig:fig2}(a) for jump operators which tend to drive the system toward the W state \cite{Dur_2000} or to  cluster states \cite{Briegel_2001}. 

However, exponential speedup is not necessary for entanglement MEs. Assuming even $N$, let us compare two initial states parametrized by angles $\phi$ and $\theta$, 
\begin{eqnarray}\nonumber
|\psi_A(\theta)\rangle&=&
\frac{|00\cdots0\rangle+\frac{\sin\theta}{4}\,|11\cdots1\rangle}{\sqrt{1+\frac{\sin^2\theta}{16}}},\\ \label{initial}
|\psi_B(\phi)\rangle&=&
\bigotimes_{m=1}^{N/2}\left(\frac{|00\rangle+0.3\sin\phi\,|11\rangle}{\sqrt{1+0.09\sin^2\phi}}\right)_m.
\end{eqnarray}
where the assignment of (L,H) labels to $(A,B)$ depends on $(\phi,\theta)$. The spectral-filtering diagrams in Figs.~\ref{fig:fig2}(b,c) show the entanglement time difference $S$ for the PEME and for the REME, respectively, see Eq.~\eqref{gain}. A relabeling discontinuity is seen for the PEME in Fig.~\ref{fig:fig2}(b), which occurs if ${\cal N}_A(0)={\cal N}_B(0)$ such that the labels H and L are exchanged without modifying asymptotic relaxation properties. The sign of $S$ then changes abruptly. For the REME in Fig.~\ref{fig:fig2}(c), in addition, a second transition due to a genuine modification of the asymptotic relaxation occurs, where long-time tails in ${\cal N}_{A,B}(t)$ coincide near the transition and $S$ smoothly vanishes. Beyond this continuous transition, the REME disappears. In addition, we find regions with exponential speedup in Fig.~\ref{fig:fig2}(c).

\emph{Conclusions.---}We have introduced different types of entanglement MEs, demonstrating an exponential speedup of entanglement generation. By identifying the spectral origin of the effect and combining it with state engineering, the origin of the speedup can be traced to the suppression of slow relaxation modes, either via the initial-state coefficients $c_n$ 
or via the Fréchet coefficients ${\cal Z}$ characterizing the entanglement monotone.
This procedure can also be applied to many-body systems, despite the existence of many slow modes. From an experimental perspective, the required initial-state rotations can be implemented through coherent control protocols including local and global qubit rotations, making entanglement speedup an accessible option on available platforms \cite{Koch2022,Lewis_2025}. To take into account the actual cost needed 
for preparing the initial state $\rho_\perp(0)$, we plan to analyze Pontus-Mpemba protocols \cite{Nava2025} for entanglement MEs in future work. 
Generalizing the present theory to multipartite entanglement \cite{Ma_2025} may enable a systematic engineering of ultrafast pathways toward preparing highly entangled complex many-body resources.

\emph{Data availability.---}The data underlying all figures in this paper will be made available on Zenodo \cite{Zenodo}.

\begin{acknowledgments} 
We thank D. Bru{\ss}, S. Diehl, D. Giuliano, Y. Gefen, and H. Kampermann for discussions. We acknowledge funding by the Deutsche Forschungsgemeinschaft (DFG, German Research Foundation) under Projektnummer 277101999 - TRR 183 (project B02), under Grants No. EG 96/14-1 and GO 1405/7-1, and under Germany's Excellence Strategy - Cluster of Excellence Matter and Light for Quantum Computing (ML4Q) EXC 2004/2 - 390534769. 
\end{acknowledgments}

\bibliography{biblio.bib}


\newpage
\appendix
\section*{End Matter}
\setcounter{equation}{0}
\setcounter{figure}{0}
\renewcommand{\theequation}{A\arabic{equation}}
\renewcommand{\thefigure}{A\arabic{figure}}

We here provide technical details on the material in the main text. To start, let us briefly address Eqs.~\eqref{REMEcond} and \eqref{PEMEcond} for degenerate or complex-valued eigenvalues $\{\lambda_n\}$.
First, let $\lambda_n$ be a degenerate eigenvalue with right eigenoperators $R_{n,\alpha}$, where $\alpha$ is a degeneracy index.  Equation \eqref{eq:rho_expansion} then contains a term 
$\left(\sum_{\alpha} c_{n,\alpha} R_{n,\alpha} \right) e^{\lambda_n t}$, and similarly $E(t)$ will contain terms like $\left(\sum_{\alpha} c_{n,\alpha} {\cal Z}_{n,\alpha} \right) e^{\lambda_n t}$ plus nonlinear contributions.
The generalization of Eqs.~\eqref{REMEcond} and \eqref{PEMEcond} is then obvious. Second, concerning complex-conjugate eigenvalues $(\lambda_n,\lambda_n^\ast)$, 
the coefficient $\tilde c_n(t)=[c_n R_n e^{i{\rm Im}(\lambda_n) t}+{\rm h.c.}]$ in front of the $e^{{\rm Re}\lambda_n t}$ factor in Eq.~\eqref{eq:rho_expansion} now becomes time-dependent. A generalization of Eqs.~\eqref{REMEcond} and \eqref{PEMEcond} follows by considering the envelopes of these coefficients.
In addition, setting $c_n=0$ eliminates the complex-conjugate mode pair both from $\rho(t)$ and $E(t)$. In the presence of exceptional points, $\delta \rho\sim t^k e^{\lambda_n t}$, polynomial corrections appear in Eqs.~\eqref{eq:rho_expansion} and \eqref{eq:E_expansion}--\eqref{PEMEcond}.

Next, we write $H$ for the I2KM in Eq.~\eqref{eq} in the computational basis
$\{|00\rangle, |01\rangle, |10\rangle, |11\rangle\}$ with $|11\rangle=c_1^\dagger c_2^\dagger|00\rangle$, which is
equivalent to two spin-$\frac12$ degrees of freedom,
$
H =\begin{pmatrix}
0 & 0 & 0 & \Delta^\ast \\
0 & \epsilon_2 & J_h & 0 \\
0 & J_h & \epsilon_1 & 0 \\
\Delta & 0 & 0 & \epsilon_1+\epsilon_2+U
\end{pmatrix}.
$
With rates $\gamma$ and $\kappa$, and using the spin-lowering operators 
$\sigma^-_{i=1,2}=\frac12(\sigma^x_i-i\sigma^y_i)$,
the set of jump operators in Eq.~\eqref{eq:lindblad} includes
\begin{equation} \label{eq:jump1}
K_1 = \sqrt{\gamma} \,\, |\Psi_{-}\rangle\langle 00|,\quad
K_2 = \sqrt{\kappa}\, \, (\sigma^-_1 + \sigma_2^-),
\end{equation}
which tend to drive the system toward the entangled state
$|\Psi_- \rangle = \frac{1}{\sqrt{2}}( |10\rangle - |01\rangle).$
With rates $\gamma_{\rm local}$ and $\gamma_\phi$, we also include jump operators describing local 
decoherence, 
\begin{equation} \label{eq:jump2}
K_3 = \sqrt{\gamma_{\mathrm{local}}}\,\, \sigma^-_1,\quad
K_4 = \sqrt{\gamma_{\mathrm{local}}}\,\, \sigma^-_2,
\end{equation}
and local dephasing,
\begin{equation} \label{eq:jump3}
K_5 = \sqrt{\gamma_{\phi}}\,\, \sigma^z_1,\quad K_6 = \sqrt{\gamma_{\phi}}\,\, \sigma^z_2.
\end{equation}
For the Mpemba protocol, two system copies are initialized in pure states $|\psi^{\rm (L)}\rangle$ and
$|\psi^{\rm (H)}\rangle$ with $E^{\rm (L)}< E^{\rm (H)}$, respectively.
For $t>0$, both copies evolve according to Eq.~\eqref{eq:lindblad} under the same parameters 
toward a common stationary state $\rho^{\rm (st)}$. 
In Fig.~\ref{fig:fig1}(a), with $J_h=1$ as energy unit, we use $U = 0.5, \Delta = 0.3, 
\epsilon_1 = \epsilon_2 = 0, \gamma = 0.5, \kappa = 0.75$, and $\gamma_{\rm local} = \gamma_\phi = 0$. 
The initial states are $|\psi^{\rm (L)}\rangle = \cos(1.05)\,|00\rangle + \sin(1.05)\,|01\rangle$ and $|\psi^{\rm (H)}\rangle= \cos(0.17)\,|01\rangle + \sin(0.17)\,|10\rangle$. In Fig.~\ref{fig:fig1}(b), we use the same Hamiltonian parameters as in Fig.~\ref{fig:fig1}(a) but with the rates $\gamma = 5$, $\kappa = \gamma_{\rm local} = 2.5$, and 
$\gamma_\phi = 2$. The pure initial states are now given by $|\psi^{\rm (L)}\rangle = -0.71\,|00\rangle-\left(0.56+0.04i\right)|01\rangle+\left(0.42+0.09i\right)|10\rangle-0.04i|11\rangle$ and $|\psi^{\rm (H)}\rangle = 0.71\,|00\rangle+\left(0.59+0.04i\right)|01\rangle-\left(0.36+0.12i\right)|10\rangle+\left(0.03+0.04i\right)|11\rangle$.

\begin{figure}[t]
\centering
\includegraphics[width=\linewidth]{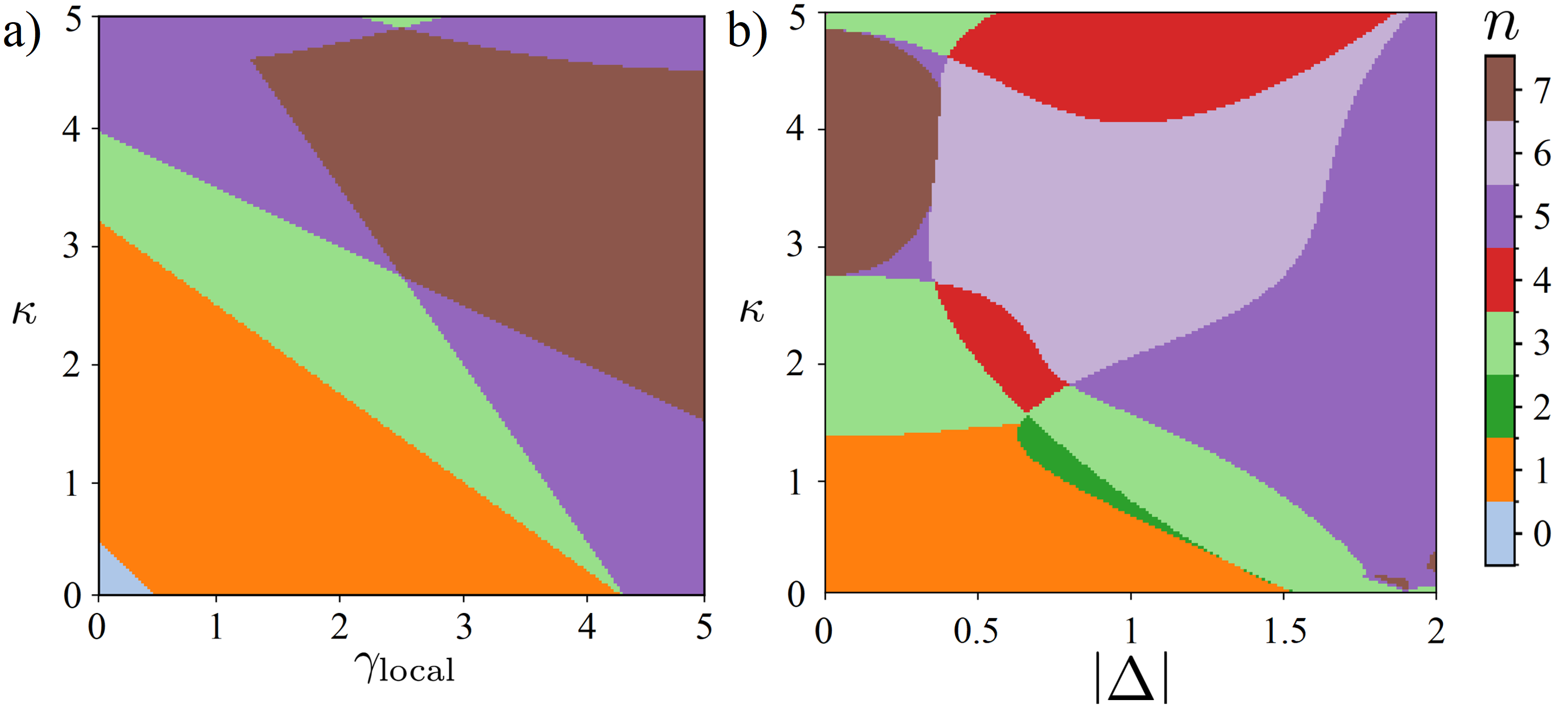}
\caption{$\mathcal{Z}_k^{({\cal N})}$ mode visibility diagram for the I2KM in the (a) $(\gamma_{\rm local},\kappa)$--plane and (b) $(\Delta,\kappa)$--plane. With $J_h=1$, we use $\Delta=0$ in (a) and $\gamma_{\rm local}=2.5$ in (b), with $U = 0.5, \epsilon_1 = \epsilon_2 = 0, \gamma = 5$, and $\gamma_\phi = 2.0$.
Each region corresponds to ${\cal Z}_k^{({\cal N})}=0$ for all $k=1,\ldots,n$ in Eq.~\eqref{eq:ZN}. Boundaries separating different regions are predominantly due to eigenvalue crossings.
}
\label{fig:figA1}
\end{figure}

We next note that by changing the Hamiltonian parameters in Eq.~\eqref{eq} and/or the rates in Eqs.~\eqref{eq:jump1}--\eqref{eq:jump3}, one can achieve $\mathcal{Z}^{(\mathcal{N})}_k=0$, see Eq.~\eqref{eq:ZN}, for all $k=1,\ldots,n$. Figure~\ref{fig:figA1} shows the corresponding mode visibility diagram in the $(\gamma_{\rm local},\kappa)$--plane and in the ($\Delta,\kappa$)--plane, where each region is characterized by the vanishing of the lowest $n$ coefficients $\mathcal{Z}^{(\mathcal{N})}_k$. The finite extent of each region highlights the robustness of the spectral filtering mechanism, such that no fine tuning of parameters is needed to obtain $\mathcal{Z}^{(\mathcal{N})}_k=0$ for $k=1,\ldots,n$. Crossing a boundary toward larger $n$ implies an exponential entanglement speedup for arbitrary initial states if $|{\rm Re}\lambda_n|<2|{\rm Re}\lambda_1|$,
see Eq.~\eqref{eq:E_expansion}, which holds for all $n$ in Fig.~\ref{fig:figA1}.

\begin{figure}
  \centering
  \includegraphics[width=0.96\linewidth]{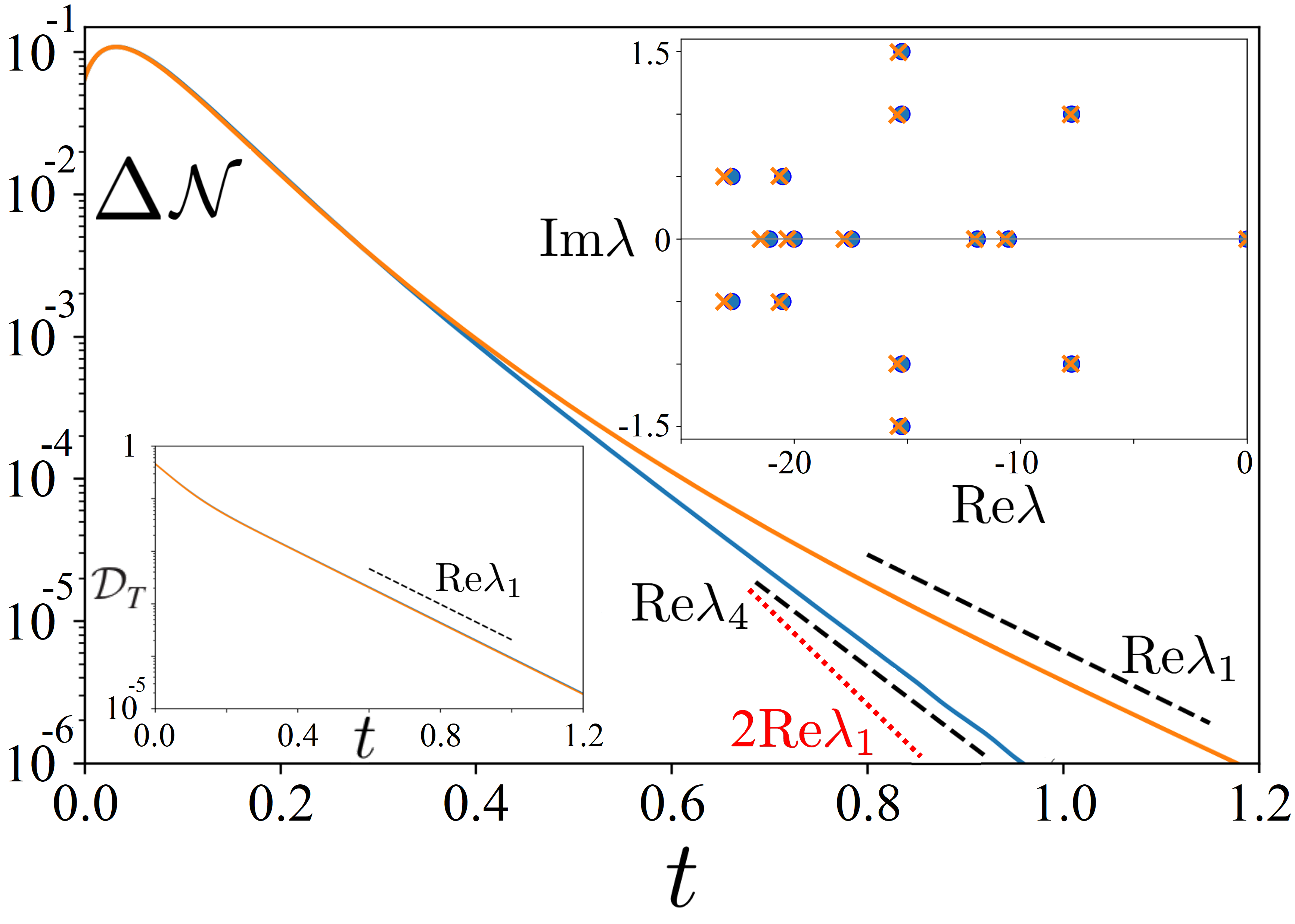}
  \caption{Logarithmic negativity difference $\Delta \mathcal{N}(t) = |\mathcal{N}(t)-\mathcal{N}_\infty|$ (shown on a logarithmic scale) vs time for the I2KM using the pure initial state $|\psi(0)\rangle = 0.71\,|00\rangle - \left(0.19+0.06i\right)|01\rangle + \left(0.63+0.06i\right)|10\rangle - \left(0.22+0.03i\right)|11\rangle$
 for two parameter sets. Their Lindbladian spectra are shown in the upper right inset. The parameter sets are very close to a boundary separating $n=0$ and $n=3$ regions. We use $J_h=1, U = 0.5, \Delta = \epsilon_1 = \epsilon_2 = 0, \gamma = 5, \kappa = 7.5, \gamma_{\rm local} = 2.5$, and $\gamma_\phi = 2$ for the blue curve ($n=3$ region). For the orange curve, we minimally perturb these parameters to enter the $n=0$ phase region; precise values are provided on Zenodo \cite{Zenodo}. 
While both parameter sets give practically identical ${\cal D}_T(t)$ curves, see bottom left inset, the $\Delta{\cal N}(t)$ curves differ drastically.}
  \label{fig:figA2}
\end{figure}

In Fig.~\ref{fig:figA2}, we address $\mathcal{N}(t)=\mathcal{N}(\rho(t))$ for the I2KM under two very similar Lindbladians (see the upper right inset),
starting from the same initial state. The corresponding trace-distance curves, see the lower left inset, are both governed by the slowest mode $(\lambda_1)$ and are essentially indistinguishable. However, we find rather different coefficients $\mathcal{Z}^{(\mathcal{N})}_k$ in Eq.~\eqref{eq:ZN} for both parameter choices, 
and therefore drastic changes in ${\cal N}(t)$.
Indeed, for the orange curve in Fig.~\ref{fig:figA2}, we have $\mathcal{Z}^{(\mathcal{N})}_1 \neq 0$, while for the blue curve, ${\cal Z}_k^{\mathcal{(N)}}=0$ with $k\in\{1,2,3\}$. Since $|{\rm Re}\lambda_4|<2|{\rm Re}\lambda_1|$ for this case, the nonlinear terms in Eq.~\eqref{eq:E_expansion} do not affect the long-time behavior of $E(t)$.
In fact, the eigenvalues can be determined analytically for the parameters of the blue curve, where the above inequality translates to ${33}/{2}-\sqrt{21} < {31}/{2}$. 
The blue curve thus converges exponentially faster to $E_\infty$ than the orange curve, 
with decay rate $\Gamma=|{\rm Re}\lambda_4|$, see Fig.~\ref{fig:figA2}. On the other hand, 
when using the concurrence ${\cal C}$ as monotone, the entanglement dynamics becomes very similar for both parameter sets. Using the concurrence vs logarithmic negativity thus makes a significant difference. 
Here we have chosen parameters resulting in an $X$-state for $\rho^{(\rm st)}$, where nonzero entries are located only on the main (anti-)diagonal \cite{Yu_2007,Mendonca_2014}. 
If also the initial state belongs to this sector, the form
$\rho(t) = \begin{pmatrix}
\rho_{11}(t) & 0 & 0 & \rho_{14}(t) \\
0 & \rho_{22}(t) & \rho_{23}(t) & 0 \\
0 & \rho_{23}^*(t) & \rho_{33}(t) & 0 \\
\rho_{14}^*(t) & 0 & 0 & \rho_{44}(t)
\end{pmatrix}$
is preserved under Eq.~\eqref{eq:lindblad} with Eqs.~\eqref{eq} and \eqref{eq:jump1}--\eqref{eq:jump3} at any time, and the concurrence reduces to $\mathcal{C}(t) =2\max(0,\mathcal{K}(t),\mathcal{Q}(t))$ with $\mathcal{K}=|\rho_{23}(t)|-\sqrt{\rho_{11}(t)\rho_{44}(t)}$ and $\mathcal{Q}(t)=
|\rho_{14}(t)|-\sqrt{\rho_{22}(t)\rho_{33}(t)}$
\cite{Wootters1998}. Assuming Fréchet differentiability, for $\mathcal{K}(t)$ we find 
\begin{equation} \label{eq:ZC}
  \mathcal{Z}^{\mathcal{(C)}}_k = 2\mathrm{Re}\left[
  \frac{\rho_{23}^{\mathrm{(st)}*}}{|\rho_{23}^{\mathrm{(st)}}|}
  (R_k)_{23} \right]-
  \frac{\rho_{44}^{\mathrm{(st)}} (R_k)_{11}+\rho_{11}^{\mathrm{(st)}} (R_k)_{44} }{\sqrt{\rho_{11}^{\mathrm{(st)}}\rho_{44}^{\mathrm{(st)}}}}.
\end{equation}
A similar formula holds for $\mathcal{Q}(t)$ by replacing $\rho_{23}\rightarrow \rho_{14}$, $\rho_{11}\rightarrow \rho_{22}$, $\rho_{44}\rightarrow \rho_{33}$, and likewise for $R_k$. 

Indeed, at long times, $\mathcal{K}(t)=\mathcal{K}( \rho^{\rm (st)} + \sum_{k\ge 1} c_k e^{\lambda_k t}R_k) \simeq \mathcal K(\rho^{\rm (st)}) +\delta\mathcal K$ with the lowest-order correction $\delta\mathcal{K}= \sum_{\alpha} \frac{\partial\mathcal K}{\partial \alpha}\Big|_{\rho^{\rm (st)}} \delta\alpha,$ 
where $\{\alpha\}$ denotes the independent real components of $\rho$. Using
$\delta |\rho_{23}| =\mathrm{Re}\!\left[  \frac{\rho_{23}^{\mathrm{(st)}\ast}}{|\rho_{23}^{\mathrm{(st)}}|} \,\delta\rho_{23}\right]$
and
$
\delta\sqrt{\rho_{11}\rho_{44}}
=\frac{\rho_{44}^{\mathrm{(st)}}\,\delta\rho_{11} + \rho_{11}^{\mathrm{(st)}}\, \delta\rho_{44}} 
{2\sqrt{\rho_{11}^{\mathrm{(st)}}\rho_{44}^{\mathrm{(st)}}}},
$
we arrive at $2\delta \mathcal{K} =\sum_{k\ge 1} c_k e^{\lambda_k t} \mathcal{Z}_k^{(\mathcal{C})}$ with $\mathcal{Z}_k^{(\mathcal{C})}$ in Eq.~\eqref{eq:ZC}.

\begin{figure}
  \centering
  \includegraphics[width=0.96\linewidth]{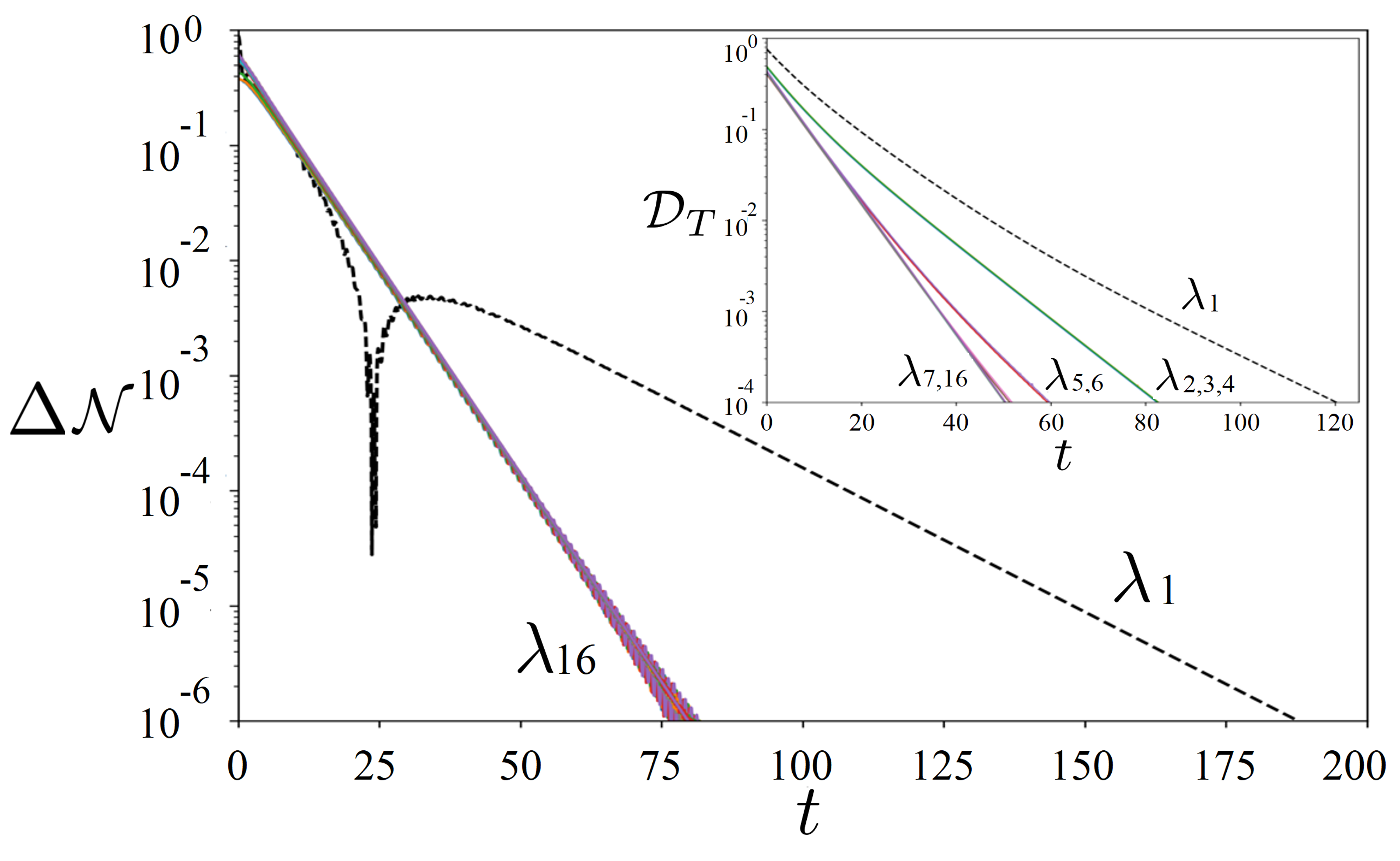}
  \caption{$\Delta \mathcal{N}(t)$ (shown on a logarithmic scale) vs time for the long-range Ising model, with parameters and $\rho(0)$ as in Fig.~\ref{fig:fig2}(a). The black dashed curve corresponds to $\mathcal{N}(\rho(t))$, solid curves to $\mathcal{N}(\rho_\perp(t))$ obtained by sequentially removing the $15$ slowest modes. Inset: Same but for ${\cal D}_T(t)$.}
  \label{fig:figA3}
\end{figure}

Next we turn to the derivation of ${\cal Z}^{\rm ({\cal N})}_k$ in Eq.~\eqref{eq:ZN}, which holds for arbitrary systems and for arbitrary stationary states.
We first introduce the negativity \cite{Vidal_2002},
$
\mathcal{E}(t)= \frac{\| \, [\rho(t)]^{T_B}\, \|_1-1}{2}
= -\sum_{\mu_j<0}\mu_j(t),
$
where $[\rho(t)]^{T_B} \, |\phi_j(t)\rangle= \mu_j(t) |\phi_j(t)\rangle.$
At long times, we can replace $|\phi_j(t)\rangle$ by the stationary eigenvectors $|\phi_j\rangle$ corresponding 
to $\rho^{\rm (st)}$, and obtain the long-time expansion 
$
\mu_j(t) \simeq \mu_j^{(\mathrm{st})} + \sum_{k\ge 1} c_k e^{\lambda_k t}
\langle\phi_j|R_k^{T_B}|\phi_j\rangle.
$
As a result, keeping only the first-order Fréchet derivative, 
$\mathcal{E}(t) \simeq \mathcal{E}^{(\mathrm{st})} + \sum_{k\ge 1} c_k e^{\lambda_k t}
{\mathcal Z}_k^{(\mathcal{E})}$ with $\mathcal Z_k^{(\mathcal{E})}
= - \sum_{\mu_j^{(\mathrm{st})}<0}
\langle \phi_j |R_k^{T_B} | \phi_j\rangle.$
Using ${\mathcal N}(t) =\log_2\!\left[1+2\mathcal{E}(t)\right]$,
we arrive at Eq.~\eqref{eq:ZN}. Similarly, for the second order, 
\[ 
\mathcal{Z}_{nm}^{(\mathcal{E})}
=\sum_{\mu_j<0}\sum_{\ell\neq j}
\frac{
\langle \phi_j | R_n^{T_B} | \phi_\ell \rangle
\langle \phi_\ell | R_m^{T_B} | \phi_j \rangle
}{\mu_\ell-\mu_j}.
\]
For the logarithmic negativity, we thereby arrive at
\begin{equation}
\mathcal Z_{nm}^{({\mathcal N})}
=\frac{2{\mathcal Z}_{nm}^{(\mathcal E)}}
{\left\|(\rho^{(\mathrm st)})^{T_B}\right\|_1 \ln 2}
-\frac{2{\mathcal Z}_n^{(\mathcal E)}
{\mathcal Z}_m^{(\mathcal E)}}
{\left\|(\rho^{(\mathrm st)})^{T_B}\right\|_1^2 \ln 2}.
\end{equation}

Finally, in Fig.~\ref{fig:figA3}, we show $\Delta \mathcal{N}(t) = |\mathcal{N}(t)-\mathcal{N}_\infty|$ for the same parameters as in Fig.~\ref{fig:fig2}(a). For an initial state with finite overlap with all slow modes, $\Delta \mathcal{N}(t)$ decays with the slowest mode ($\lambda_1$). By progressively eliminating the $15$ slowest modes, 
we observe that the resulting $\Delta{\cal N}(t)$ curves cluster together, i.e., they share the same asymptotic decay law. In contrast, the trace-distance dynamics, see inset of Fig.~\ref{fig:figA3}, shows distinct small-size clusters.

\end{document}